\documentclass[a4paper,aps,prl,twocolumn,10pt, 
superscriptaddress,floatfix, showpacs]{revtex4-1}

\usepackage[utf8]{inputenc}
\usepackage[english]{babel}

\usepackage{amsmath,amssymb,amsfonts}
\usepackage{graphicx}
\usepackage{color}
\usepackage{units}

\usepackage{mathrsfs}

\usepackage{pdfcomment}

\newcommand{\up}{\uparrow}
\newcommand{\down}{\downarrow}

\newcommand{\Tr}{\textrm{Tr}}
\newcommand{\dd}[1]{\mathrm{d}#1\,}

\newcommand{\kb}{k_\mathrm{B}}

\newcommand{\calF}{\mathcal{F}}
\newcommand{\calG}{\mathcal{G}}
\newcommand{\calK}{\mathcal{K}}

\newcommand{\sgn}{\mathop{\mathrm{sgn}}}

\begin{document}
 
\title{Controlling spin polarization of a quantum dot via a helical edge state}
\author{Benedikt Probst}
\affiliation{Institut für Mathematische Physik, TU Braunschweig, Mendelssohnstr. 
3, 38106 Braunschweig, Germany}
\author{Pauli Virtanen}
\affiliation{O.V. Lounasmaa Laboratory, Aalto University,
  P.O. Box 15100, FI-00076 AALTO, Finland}
\author{Patrik Recher}
\affiliation{Institut für Mathematische Physik, TU Braunschweig, Mendelssohnstr. 
3, 38106 Braunschweig, Germany}
\affiliation{Laboratory for Emerging Nanometrology Braunschweig, 38106 
Braunschweig, Germany}
\date{\today}

\begin{abstract}
We investigate a Zeeman-split quantum dot (QD) containing a single spin $1/2$ 
weakly coupled to a helical Luttinger liquid (HLL) within a generalized master 
equation approach. The HLL induces a tunable magnetization direction on the QD 
controlled by an applied bias voltage when the quantization axes of the QD and 
the HLL are noncollinear. The backscattering conductance (BSC) in the HLL is 
finite and shows a resonance feature when the bias voltage equals the Zeeman 
energy in magnitude. The observed BSC asymmetry in bias voltage directly 
reflects the quantization axis of the HLL spin.
\end{abstract}


\maketitle

The hallmark of time reversal invariant topological insulators (TI) 
\cite{Hasan2010,Qi2011} in two dimensions is the quantum spin Hall (QSH) effect. 
The edge states forming at the boundary of the QSH device are counterpropagating 
Kramers pairs. The QSH effect was proposed and measured in HgTe/CdTe 
\cite{Bernevig2006,Konig2007,Roth2009} and InAs/GaSb quantum wells (QWs) 
\cite{Liu2008,Knez2011}. The spin polarization in the edge state transport has 
been demonstrated by combining the QSH effect and the spin Hall effect 
\cite{Brune2012}. The QSH edge states form a helical Luttinger liquid (HLL) 
\cite{Wu2006} in the presence of electron-electron interactions.

The helical structure imposes strong restrictions for backscattering, and allows 
many mechanisms to be potentially important. Effects of single spin impurities 
coupled to a HLL for isotropic \cite{Tanaka2011,Maciejko2009,Schiller1995,Vayrynen2014} and 
anisotropic Kondo models \cite{Eriksson2012,Eriksson2013a,Vayrynen2014}, as well 
as in tight-binding models for graphene ribbons with spin orbit interaction 
within the Kane--Mele model \cite{Narayan2013,Hurley2013,Hu2013,Goth2013} have 
been considered. In a similar context, the effect of backscattering by nuclear 
background spins \cite{Lunde2012} and the implications of this background for 
Rashba scattering \cite{DelMaestro2013} have also been studied. In all these 
works, no dc backscattering current is found without intrinsic spin relaxation, 
an anisotropic Kondo coupling or Rashba impurity, due to the necessity to flip 
the spin in order to scatter between opposite branches of the QSH edge. This 
characteristic is reminiscent of QDs coupled to bulk (3D) ferromagnetic leads, 
where the coupling to the leads controls the QD behavior, and virtual exchange 
of electrons induces an exchange field on the QD parallel to the lead 
polarization \cite{Braun2004,Weymann2005,Weymann2006} adding to possible 
external fields \cite{Gorelik2005,Urban2007}.  

\begin{figure}
 \includegraphics[width=\columnwidth]{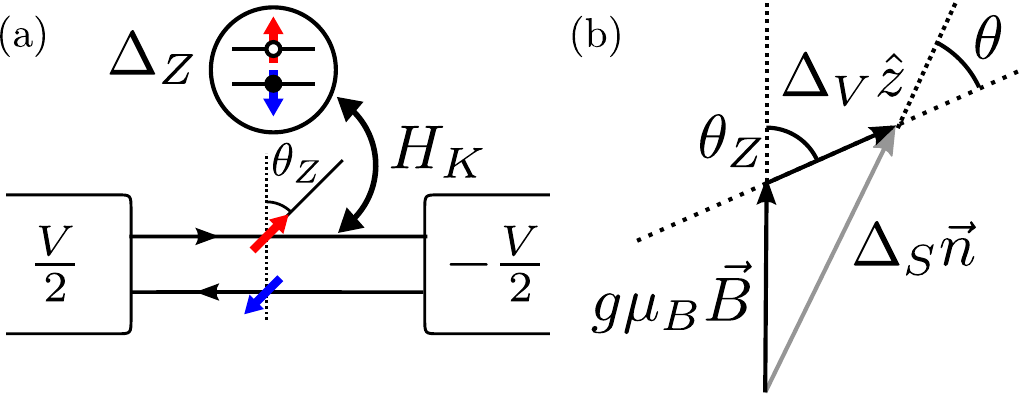}
 \caption{ Setup considered in this paper (a) and construction of the effective 
QD Hamiltonian (b). A HLL is coupled to a QD described by a Kondo Hamiltonian 
$H_K$. A magnetic field $\vec B$ is applied to the QD, which is tilted with 
respect to the quantization axis $\hat z$ of the helical edge state (fixed by 
the helical edge) by an angle $\theta_Z$. The effective system Hamiltonian for 
the QD is the sum of a Zeeman term $g\mu_B\vec B\cdot\vec S$ and an induced part 
$\Delta_V S^z$ that corresponds to the spin-polarization of the HLL driven by a 
bias voltage $V$. The resulting effective field points along $\vec n$ with 
tilt-angle $\theta$ and has strength $\Delta_S$ (see text).
 }
 \label{fig:setup}
\end{figure}

In this work, we discuss a setup in which a QD in the cotunneling regime is 
coupled to the QSH device, which acts as a spin polarized reservoir. In order to 
manipulate the spin on the QD, we assume that a magnetic field is applied to the 
QD, inducing a quantization axis that is not parallel with that of the QSH 
states that is determined by spin-orbit interaction. The resulting dynamics are 
described with a general master equation (GME) approach. From the GME we obtain 
the spin polarization of the QD and the transport signatures, including effects 
due to electron-electron interactions in the helical edge, which is crucial for 
these 1D channels. We discuss the possibilities to manipulate the state of the 
QD, and the signatures of these manipulations in the transport properties.

{\it Model.--} We consider a Zeeman split QD in the Coulomb blockade regime 
coupled to a HLL, as illustrated in Fig.~\ref{fig:setup}(a). The Hamiltonian is 
given by (we set $\hbar=e=1$)
\begin{equation}
  H=H_\mathrm{HLL}+H_Z+H_K.
\end{equation}
The HLL is described by the Hamiltonian~\cite{Wu2006}
\begin{align}
  H_\mathrm{HLL}
  &=v_F\int\dd{\xi}:\sum_{\eta=\pm}
\Psi_\eta^\dagger(-i\eta\partial_\xi)\Psi_\eta(\xi):
  \\\notag
  &\quad+\frac{\lambda}{2}\int 
\dd{\xi}:\Big(\sum_{\eta=\pm}\Psi_\eta^\dagger(\xi)\Psi_\eta(\xi)\Big)^2:,
  \label{eq:HBasic} 
\end{align}
where $v_{F}$ is the Fermi velocity, $\Psi^{(\dagger)}_\eta$ are electron 
operators on the edge in branch $\eta=\pm$ corresponding to right and left 
movers, which are spin polarized parallel or antiparallel to the spin 
quantization axis $\hat z$ due to the helical nature of the HLL, and $\lambda$ 
is the interaction strength due to Coulomb repulsion in the helical edge. The 
QSH device is connected to leads and a bias $V$ is applied.  As the electrons 
have a definite propagation direction correlated with spin, this induces a spin 
bias. The bias can be gauged into the lead operators~\cite{Peca2003}.

In the cotunnelling regime, a single-level QD coupled to a HLL can be described 
by the Kondo Hamiltonian \cite{Hewson1993}
\begin{equation}
  \label{eq:Hkondo}
  H_K=J\big(J^+S^-+J^-S^++2J^zS^z\big),
\end{equation}
where $J$ is the Kondo coupling strength.  The spin operators are defined by 
$J^\tau=\tfrac{1}{2}\sum_{\mu,\nu}\Psi_\nu^\dagger(0)\sigma^\tau_{\nu,\mu} 
\Psi_\mu(0)$ and 
$S^\tau=\tfrac{1}{2}\sum_{\mu,\nu}d_\nu^\dagger\sigma^\tau_{\nu,\mu}d_\mu$, 
where $\sigma^\tau$, $\tau=x,y,z$, are the Pauli matrices and where 
$d^{(\dagger)}_\sigma$ correspond to electrons with spin projection $\sigma$ on 
the QD with respect to $\hat z$.  The ladder-operators are defined as $J^\pm=J^x\pm 
i J^y$ and $S^\pm=S^x\pm i S^y$.

We allow for a general direction of the magnetic field $\vec B$ on the QD, and 
the Zeeman term obtains its standard form,
\begin{equation}
  \label{eq:Hzeeman}
  H_Z = g\mu_B \vec B\cdot\vec S
  ,
\end{equation}
where $g$ is the $g$-factor, $\mu_B$ is the Bohr magneton, and 
$\Delta_Z=g\mu_B|\vec{B}|$ the resulting Zeeman splitting \cite{Note1}.

{\it Spin dynamics for QD.--} We consider the dynamics of the QD spin in the 
weak coupling limit ($J\to0$) and derive a generalized master equation 
(GME)~\cite{Koller2010,Breuer2002,Blum1996} for the reduced density matrix 
$\rho\equiv\Tr_B\rho_\mathrm{tot}$ of the QD, where the degrees of freedom of 
the bath (the HLL) are traced out. For convenience, we redefine the system, 
bath, and interaction Hamiltonian as $H_S\equiv H_Z + 2J \langle {J^z}\rangle_0 
{S^z}$, $H_B\equiv H_\mathrm{HLL}$, and $H_I\equiv H_K - 2J \langle 
{J^z}\rangle_0 {S^z} =\sum_{k=\pm,z}A_kS^k$, where the nonvanishing expectation 
value of $H_K$  is absorbed into $H_S$ \cite{Note2}.  Indeed, due to the 
finite bias $V$, we have $\langle{J^z}\rangle_0=V/(8\pi v_F)$, where 
$\langle{\ldots}\rangle_0$ denotes expectation values for the uncoupled system 
($H_I=0$). Here, we also define $A_\pm\equiv J {J^\mp}$ and $A_z\equiv 2 J 
({J^z}-\langle {J^z}\rangle_0)$.  $H_S$ describes the interaction with the 
external magnetic field and an effective HLL-induced one ($\propto JV$), as 
illustrated in Fig.~\ref{fig:setup}(b).  The total effective field is in the 
plane spanned by the HLL spin-quantization axis and the external field.

The effective magnetic field felt by the QD can be parametrized using polar 
coordinates $\theta$ and $\phi$. Defining $\Delta_V\equiv{}2J\langle 
{J^z}\rangle_{0}$, the system Hamiltonian can be written as \begin{equation} 
H_S=\Delta_S{\vec n}\cdot\vec S, \end{equation} where  
$\Delta_S=(\Delta_V^2+\Delta_Z^2+2\Delta_V\Delta_Z\cos\theta_Z)^{1/2}$ and 
${\vec n}=(\sin\theta\cos\phi, \sin\theta\sin\phi, \cos\theta)^{\rm T}$ is the 
direction of the effective magnetic field. The polar angle for this effective 
field is defined as 
$\theta\equiv\arccos[(\Delta_Z\cos\theta_Z+\Delta_V)/\Delta_S]$, where 
$\theta_Z$ is the angle between $\vec B$  and lead quantization axis $\hat z$.

This Hamiltonian can be diagonalized by a unitary transformation 
$UH_SU^{\dagger}=\Delta_S{S'}^{z}$ where ${S'}^{z}$ is the spin along ${\vec 
n}$. This transformation corresponds to a rotation from the $\vec S$-operators 
to ${\vec S}'$ -operators defined by 
$S^{i}=\sum_{ij}\mathscr{D}(U)_{ij}{{S'}^j}$ where $i,j=z,+,-$ and 
\begin{equation}
  \label{eq:OrthogonalRot}
  \mathscr{D}(U)=
    \begin{pmatrix}
\cos\theta&-\tfrac{z_\gamma}{2}\sin\theta&-\tfrac{z_\gamma^*}{2}\sin\theta\\
      z_\phi\sin\theta&z_\phi 
z_\gamma\cos^2\frac{\theta}{2}&-z_\gamma^*z_\phi\sin^2\frac{\theta}{2}\\
      z_\phi^*\sin\theta&-z_\gamma 
z_\phi^*\sin^2\frac{\theta}{2}&z_\gamma^*z_\phi^*\cos^2\frac{\theta}{2} 
      \end{pmatrix}.
\end{equation}
Here,  $z_\phi\equiv e^{i\phi}$ and $z_\gamma\equiv e^{i\gamma}$. As $H_S$  is 
invariant under rotation around ${\vec n}$,  a free parameter $\gamma$ remains 
that will not influence the results in the lab-frame (unprimed-frame). For 
definiteness, we choose $\phi=\pi/2$ in the following so that ${\vec n}$ lies in 
the $y$-$z$-plane.

{\it Generalized master equation.--} Expanding up to second order in $H_I$ 
following standard steps \cite{Blum1996}, we arrive at the Redfield master 
equation,
\begin{subequations}
\label{eq:RedfieldMaster}
 \begin{equation}
  \dot{\rho}^I(t)=-\int_{t_0}^t\dd{\tau} 
\mathcal{K}^I(t,\tau)\lbrace\rho^I(\tau)\rbrace
 \end{equation}
 \begin{multline}
\calK^I(t,\tau)\lbrace\rho^I(\tau)\rbrace=\sum_{k,l=\pm,z}e^{
i\Delta_S(\sigma_k\tau+\sigma_l 
t)}\\
  \times\left(\calG_{k,l}(t-\tau)[{{S'}^k},{S'}^l\rho^I(\tau)]\right.\\
  +\left. \calG_{l,k}(\tau-t)[\rho^I(\tau){S'}^l,{S'}^k]\right),
 \end{multline}
\end{subequations}
where ${S^k}'$ are the spin operators defined via Eq.~\eqref{eq:OrthogonalRot}, 
$\sigma_\pm=\pm1$, $\sigma_z=0$ and 
$\calG_{k,l}(\tau)\equiv\sum_{\alpha=\pm,z}c_{\alpha,k}c_{\bar\alpha,l}G_{\alpha 
,\bar\alpha}(\tau)$, $G_{\alpha,\bar\alpha}(\tau)\equiv\langle 
A_\alpha(\tau)A_{\bar\alpha}(0)\rangle$, and 
$c_{k,l}\equiv(\mathscr{D}(U))_{k,l}$.  The bar denotes interchange of $+$ and 
$-$, with $\bar{z}=z$.  Assuming fast decaying bath correlation functions, the 
Markov approximation $\rho(t)\approx\rho(\tau)$ can be used. The terms for which 
$\sigma_k+\sigma_l\ne0$ have a coefficient whose phase oscillates with $\Delta_S 
t$ whereas the relaxation time is proportional to $\gamma_R^{-1}$ where 
$\gamma_R=(J^2/v_F^2\beta)$ with $\beta=1/k_BT$ (see 
Eqs.~(\ref{eq:Fouriertransf})). For $\gamma_R\ll \Delta_S$, those terms can be 
neglected.  Therefore, in the regime $\gamma_R\ll\Delta_S,k_BT$, the GME is 
given in Lindblad form
\begin{subequations}
  \begin{align}
    \dot\rho^I(t)&=-i[H_\mathrm{LS},\rho^I(t)]+\mathcal{D}(\rho^I(t))\\
    \mathcal{D}(\rho^I(t))&=\sum_{k,l=\pm,z}\calF_{\bar k, 
k}(-\Delta_S\sigma_k)\nonumber\\    
&\times\Big({{S'}^k\rho^I(t){{S'}^k}^\dagger}-\frac{1}{2}\lbrace{{S'}^k}
^\dagger{S'}^{k},\rho^I(t)\rbrace\Big),
  \end{align}
\end{subequations}
where $\calF_{k,l}(\omega)\equiv\int_{-\infty}^\infty \dd{\tau} 
e^{i\omega\tau}\calG_{k,l}(\tau)=\sum_{\alpha=\pm,z}c_{\alpha,k}c_{\bar\alpha,l} 
F_{\alpha,\bar\alpha}(\omega)$ are Fourier transforms of the bath correlation 
functions, and $H_\mathrm{LS}$ is a Lamb shift Hamiltonian. The latter does not 
influence the steady state, as it only leads to phase oscillations of the 
off-diagonal entries, which decay due to dephasing. The bath correlation 
functions can be found using a standard approach \cite{Giamarchi2007},
\begin{subequations}
  \label{eq:Fouriertransf}
  \begin{align}    
F_{zz}(\omega)&=\gamma_R\frac{K}{2\pi}\frac{\omega\beta}{\sinh(\omega\beta/2)}e^
{\frac{\omega\beta}{2}}\\
    F_{\pm,\mp}(\omega)&=\gamma_R\frac{(2a)^{2K-2} \; 
e^{\frac{\beta}{2}(\omega\mp V)}}{\Gamma(2K)|\Gamma(1-K-i(\omega\mp 
V)\beta/2\pi)|^2}\nonumber\\
    &\times\frac{K^2\pi}{\cosh((\omega\mp V)\beta)-\cos(2\pi K)},
  \end{align}
\end{subequations}
where $a\equiv K\pi\alpha/\beta v_F$ with $\alpha$ the short-distance cutoff and 
$K\equiv 1/\sqrt{(1+\lambda/\pi v_F)}$ is the Luttinger liquid parameter. The 
resulting steady state is
\begin{equation}
\label{steadystate}
 \bar\rho=\frac{1}{\calF_{+-}(\Delta_S)+\calF_{-+}(-\Delta_S)}
 \begin{pmatrix}
  \calF_{-+}(-\Delta_S)
  &
  0
  \\
  0
  &
  \calF_{+-}(\Delta_S)
 \end{pmatrix}
 \,.
\end{equation}
Note that  $\bar\rho$ is diagonal only in the eigenbasis of $H_S$.

\begin{figure}
\begin{center}
 \includegraphics[width=\columnwidth]{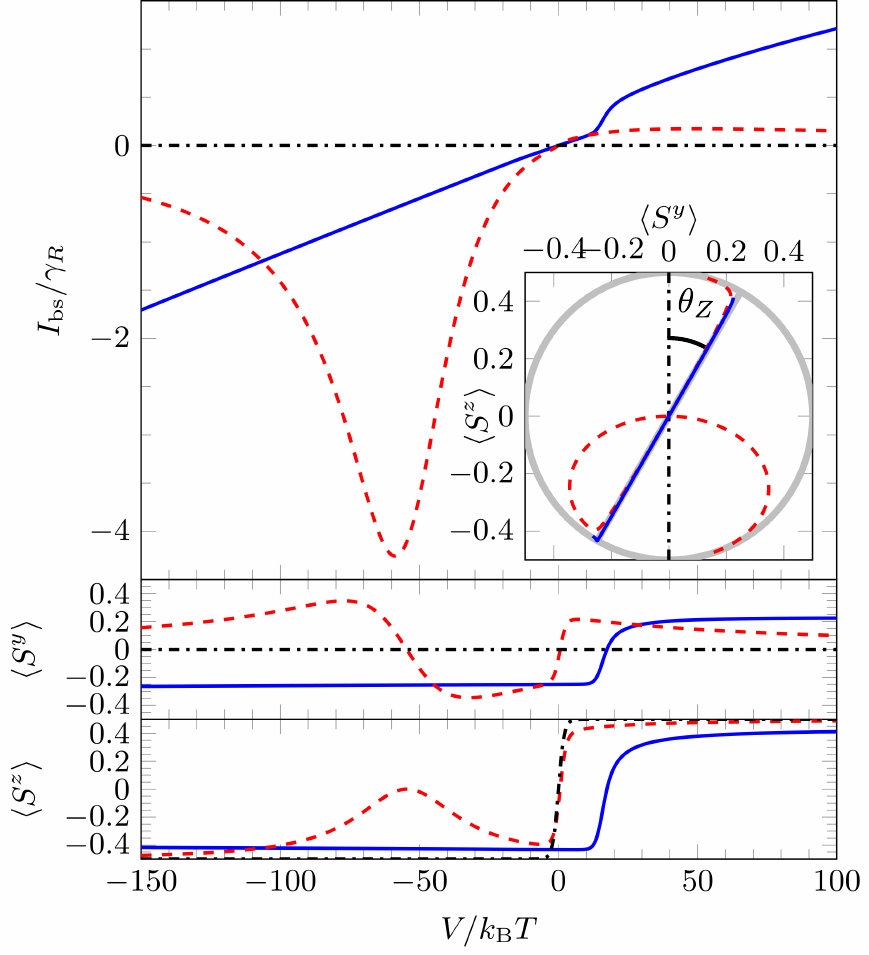}
\end{center}
\vspace{-0.8cm}
\caption{ Backscattering current and QD spin-polarization (SP) in $y$ and $z$ 
direction for $J/v_F=0.1$, $\alpha /\beta v_F=10^{-3}$, $\theta_Z=\pi/6$, 
$\Delta_Z=15~\kb T$ (blue solid), $\Delta_Z=0.5~\kb T$ (red dashed) and 
$\Delta_Z=0$ (black dash dotted) as a function of the bias $V$ and for $K=1$ is 
shown. The inset shows the SP in the $z$-$y$-plane for the same bias range. For 
illustration, the Bloch sphere and the direction of the Zeeman field (with angle 
$\theta_Z$) are added in gray. When $|V|$ is comparable to $\Delta_Z$ the SP is 
along ${\vec B}$ whereas it is along ${\hat z}$ (the HLL quantization axis) when 
$|\Delta_V|$ is larger than $\Delta_Z$. The spin flip along the magnetic field 
direction and the peculiar change of the SP within the $z$-$y$-plane shows clear 
signatures in the backscattering current.
    }
   \label{fig:CurrPol2D}
\end{figure}

{\it QD spin-polarization.--} As the effective magnetic field along ${\vec n}$ 
is tilted with respect to the Zeeman field ${\vec B}$ and the lead quantization 
axis, the QD spin-polarization (SP) in $z$ and $y$ direction $\langle 
S^{z,y}\rangle={\rm Tr}[ {\bar \rho}\,S^{z,y}]$ can deviate from both 
directions.  In order to understand the effect of this tilt, the SP as a 
function of bias $V$ for several $\Delta_Z$ and a fixed tilt $\theta_Z$ is shown 
in the inset of Fig.~\ref{fig:CurrPol2D}.  Three different regimes with respect 
to $\Delta_Z/\Delta_V$ can be distinguished.  If $\Delta_Z$  vanishes, the 
system quantization axis ${\vec n}$, and hence the SP, is aligned or 
anti-aligned with the HLL quantization axis $(\theta=0, \pi)$ , depending  on 
the sign of $V$ (dashed-dotted line in Fig.~\ref{fig:CurrPol2D}). If $\Delta_Z$ 
is finite, two further regimes can be distinguished. If $|\Delta_V|\ll \Delta_Z$ 
(full line in Fig.~\ref{fig:CurrPol2D}), the quantization axis of the QD is 
fixed to the direction of ${\vec B}$. For $|V|,k_B T\ll \Delta_Z$, the QD spin 
stays in the ground state. When  $V \approx \Delta_Z$, the QD spin can flip and 
eventually will occupy mostly the excited state. When $-\Delta_V\sgn({\hat 
z}\cdot {\vec B})\approx \Delta_Z$ there is a strong bias dependence of 
$\theta$. The QD depolarizes ($\langle S^{z}\rangle=\langle S^{y}\rangle=0$) if 
$V\cos\theta=\Delta_S$, which can be reached for positive and negative bias 
voltages, as illustrated by the dashed line in Fig.~\ref{fig:CurrPol2D}. This 
feature is a direct consequence of the helical nature of the edge states.

\begin{figure}
\begin{center}
 \includegraphics[width=\columnwidth]{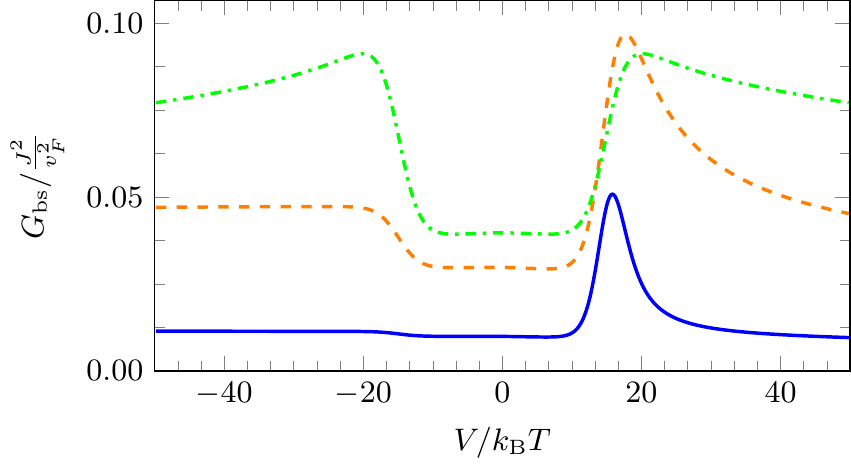}
\end{center}
\vspace{-0.8cm}
  \caption{ Differential conductance $G_{\rm bs}$ as a function of bias voltage 
$V$ for different tilt-angles: $\theta_Z=\pi/2$ (green dash dotted), 
$\theta_Z=\pi/3$ (orange dashed) and $\theta_Z=\pi/6$ (blue solid).  
$\Delta_Z=15~\kb T$, $J/v_F=0.1$, $\alpha/\beta v_F=10^{-3}$ and $K=1$. $G_{\rm 
bs}$ shows a peak when $|V|\approx\Delta_Z$ with an asymmetry between $V>0$ and 
$V<0$ that vanishes when the Zeeman field is perpendicular to the lead 
quantization axis ($\theta_Z=\pi/2$).}
\label{fig:DiffCondsec}
\end{figure}

\begin{figure}
\begin{center}
 \includegraphics[width=\columnwidth]{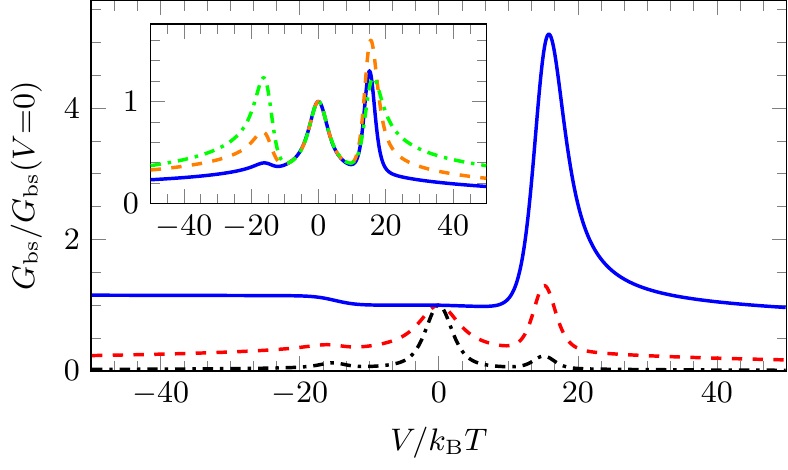}
\end{center}
\vspace{-0.8cm}
\caption{ Renormalized differential conductance as a function of bias voltage 
$V$ for different HLL interaction strengths: $K=0.6$ (black dash dotted), 
$K=0.8$ (red dashed) and $K=1$ (blue solid). $\Delta_Z=15~\kb T$, $J/v_F=0.1$, 
$\alpha/\beta v_F=10^{-3}$, and $\theta_Z=\pi/6$. The inset shows the 
renormalized differential conductance for the parameters from 
Fig.~\ref{fig:DiffCondsec} with $K=0.8$. }
\label{fig:DiffCondInteraction}
\end{figure}

{\it HLL backscattering current.--} The current correction due to the coupling 
to the QD in one of the $\eta$-branches of the HLL is given by $\hat 
I_\eta(t)=\dot N_\eta$, with
\begin{equation}
\dot N_\eta=i[H_I(t),N_\eta(t)]=i\eta(A_+(t)S^+(t)-A_-(t)S^-(t)),
\end{equation}
where $N_{\eta}=\int d\xi \Psi^{\dagger}_{\eta}(\xi)\Psi_{\eta}(\xi)$. The 
steady state current can be obtained from 
$I_\mathrm{bs}=\lim_{t\rightarrow\infty} I_-(t)=\lim_{z\rightarrow 0^-}zI_-(z)$, 
where $I_-(z)$ is the Laplace transform of $\Tr(\rho(t)\hat I_-(t))$. The 
current can now be obtained from the Born approximation for $\rho$. As the 
steady state density matrix is diagonal in the eigenbasis of $H_S$ we find for 
the backscattering current
\begin{multline} 
I_\mathrm{bs}=\big[\cos^4\frac{\theta}{2}\big(F_{-+}(-\Delta_S)\bar\rho_{\down,
\down}-F_{+-}(\Delta_S)\bar\rho_{\up,\up}\big)\\ 
+\sin^4\frac{\theta}{2}\big(F_{-+}(\Delta_S)\bar\rho_{\up,\up}-F_{+-}
(-\Delta_S)\bar\rho_{\down,\down}\big)\\
 +\frac{\sin^2\theta}{4}\big(F_{-+}(0)-F_{+-}(0)\big)
 \big].\label{eq:Currsec}
\end{multline}
The rich backscattering transport characteristics of this system is shown in 
Figs.~\ref{fig:CurrPol2D}, \ref{fig:DiffCondsec} and 
\ref{fig:DiffCondInteraction}. The backscattering current vanishes identically 
in the case $\theta=0,\pi$ (dashed-dotted line Fig.~\ref{fig:CurrPol2D}) which 
holds for ${\vec B}=0$ or if ${\vec B}$ is parallel or antiparallel to the HLL 
quantization axis (${\hat z}$), which can be seen explicitly from the first two 
lines in  Eq.~(\ref{eq:Currsec}) noting that 
$\bar\rho_{\up,\up}/\bar\rho_{\down,\down}=F_{-+}(-\Delta_S)/F_{+-}(\Delta_S)$  
in that case. This is consistent with the results given for the Kondo impurity 
in a HLL \cite{Tanaka2011,Vayrynen2014}, the isotropic Kondo impurity with Rashba interaction 
\cite{Eriksson2012}, and a QD in the Coulomb blockade regime connected to fully 
polarized antiparallel ferromagnetic leads \cite{Weymann2005a}.

However, we find that the BSC is in general finite for non-zero angle $\theta$. 
We first consider the non-interacting case $K=1$. We start with the regime 
$|\Delta_V|\ll \Delta_Z$ where $\theta\approx \theta_Z$, displayed for the 
backscattering current in Fig.~\ref{fig:CurrPol2D} (full line) as a function of 
bias voltage $V$ for a relatively small tilting angle $\theta_Z=\pi/6$ and in 
Fig.~\ref{fig:DiffCondsec} for the BSC for various angles $\theta_Z$. The clear 
conductance peak at $V\approx \Delta_Z$ reflects the energy threshold to flip 
the spin $\downarrow\rightarrow \uparrow$ on the QD, releasing the bottleneck 
for backscattering processes. For a finite angle $\theta_Z$, the second line in 
Eq.~(\ref{eq:Currsec}) starts to contribute which leads to a finite dc-BSC. 
Increasing the angle $\theta_Z$ further results in a development of a mirror 
peak at $V\approx -\Delta_Z$ since now the QD spin has an appreciable overlap 
with both spin-directions in the HLL (see Fig.~\ref{fig:DiffCondsec}). In 
addition, a constant shift of the BSC originates from the third line of 
Eq.~(\ref{eq:Currsec}) which describes spin flip processes in the leads, but 
none on the QD, and is therefore independent of ${\bar \rho}$. Turning on 
HLL-interactions, an additional peak at zero bias, cut by temperature,  starts 
to develop as well (see Fig.~\ref{fig:DiffCondInteraction}). However, the peaks 
at finite bias are still visible.

We now turn to the regime $-\Delta_V\sgn({\hat z}\cdot {\vec B})\approx 
\Delta_Z$, 
where $\theta$ strongly depends on the bias voltage $V$. The backscattering 
current in this regime is governed by a second peak as shown in 
Fig.~\ref{fig:CurrPol2D} by the dashed line and reflects the sweep of the angle 
$\theta$ through $\pi/2$ as a function of $V$ whereas the fixed angle 
$\theta_Z=\pi/6$ is rather small. In accordance with the polarization plot in 
Fig.~\ref{fig:CurrPol2D}, the second current peak appears close to the 
depolarization of the QD ($\langle S_z\rangle=\langle S_y\rangle=0$) at finite 
negative bias voltage where $\theta\approx\pi/2$. At very large $|V|$, the 
current 
has to disappear as the effective quantization axis of the QD aligns ($V>0$) or 
anti-aligns ($V<0$) with the quantization axis of the HLL.

{\it Experimental feasibility.--} Limiting the effect of ${\vec B}$ to the QD 
might be challenging in a real experiment, as e.g. a homogenous field influences 
also the helical edge. However, the orbital effect of the magnetic field 
\cite{Note3} is not expected to destroy the helical property up to several Tesla 
in devices made from HgTe/CdTe QWs \cite{Tkachov2010, Tkachov2011, Scharf2012, 
Pikulin2014,Pikulin2014}. The desired Zeeman splitting $\Delta_Z$ is smaller 
than the voltage range we consider, which in turn is much smaller than the bulk 
band gap $2D$. Assuming that $g$-factors for QD and helical edge are similar, 
the gap opened in the helical edge by a spin-mixing field will not exceed the 
band gap. A gap will open at the Dirac point and mix the two spins there, but 
for momenta far away from that point the helical behavior is maintained by the 
spin-orbit interaction \cite{Note4}. For HgTe/CdTe QWs $D$ is 
approximately $\unit[10]{meV}$. For a short-distance cutoff $\alpha\approx v_F/D$, the 
figures shown here correspond to a temperature of $T\approx \unit[116]{mK}$, 
which is experimentally feasible. For the parameters shown here the Kondo 
temperature $T_K \approx (D/k_B)\exp(- \pi v_F/J) \ll T $. 
The backscattering current in Fig.~\ref{fig:CurrPol2D} is then given in units of 
$16~\mathrm{pA}$. Similar arguments would apply to InAs/GaSb QWs. Novel proposed 
QSH materials with much larger bulk gaps 
\cite{Zhou2014,Zhang2014,Xu2013,Xiao2011} would allow for higher temperatures 
and hence energies, which would lead to larger transport signals.

In summary, we discuss the dynamics of a QD spin subject to a weak external 
Zeeman field coupled to a QSH edge state. The QD spin-polarization is determined 
by the sum of this external field and an effective field induced by the helical 
edge state spin bias. The transport signatures show a strong dependence on the 
relative orientation of magnetic field and quantization axis in the helical edge 
state. These effects enable electrical control of the QD spin, and the 
determination of the edge state spin direction. The knowledge of the spin 
direction, and its variation with energy \cite{Virtanen2012,Schmidt2012}, is 
crucial for understanding inelastic backscattering mechanisms and for 
spintronics applications. 

BP and PR acknowledge financial support from the NTH school for contacts in 
nanosystems and the EU-FP7 project SE2ND. PV acknowledges financial support by 
the Academy of Finland. We acknowledge insightful discussions with W.A. Coish, 
E.M. Hankiewicz and A. Ström. PR acknowledges the hospitality of the Tokyo 
Institute of Technology, Japan, where part of the work has been carried out. 

%

%

\end{document}